\documentclass{mn2e}
\usepackage{amsmath}
\usepackage{amssymb}
\usepackage{amsfonts}
\usepackage{graphicx}
\newcounter{tr}
\newcommand{\va}{\vec{A}}
\newcommand{\vb}{\vec{B}}
\renewcommand{\l}{\lambda}

\newcommand{\sfr}{^{\frac{1}{2}}}

\newcommand{\bqe}{\begin{eqnarray}}
\newcommand{\eqe}{\end{eqnarray}}
\title[Stability of Magnetic Equilibria in  Radio Bubbles]{Stability 
of Magnetic Equilibria in Radio Bubbles}
\author[Gregory Benford]{Gregory Benford \thanks{email:gbenford@uci.edu}\\
Department of Physics and Astronomy\\
University of California\\
Irvine, CA 92697-4575, USA}
\begin{document}
\date{\today}
\maketitle
\label{firstpage}
\begin{abstract}
Current-carrying flows, in the laboratory and in astrophysical
jets, can form remarkably stable magnetic structures. Decades of 
experience shows that such flows often build equilibria that reverse 
field directions, evolving to an MHD Taylor state, which has 
remarkable stability properties. We  model jets and the magnetic 
bubbles they build as reversed field pinch equilibria by assuming
the driver current to be stiff in the MHD sense.
Taking the jet current as rigid and a fixed function of position, we 
prove a theorem: that the same, simple MHD stability
conditions guarantee stability, even after the jet turns off.  This 
means that magnetic
structures harboring a massive inventory of
magnetic energy can persist long after the building jet current
has died away. These may be the relic radio ``fossils," ``ghost 
bubbles" or ``magnetic balloons" found in clusters. These equilibria 
under magnetic
tension will evolve, retaining the stability properties from that 
state. The remaining fossil is not a disordered ball of magnetic 
fields, but a stable
structure under tension, able to respond to the slings and arrows of 
outside forces. Typically their Alfven speeds greatly exceed the 
cluster sound speed, and so can keep out hot cluster plasmas, leading 
to x-ray ``ghosts."
Passing shocks cannot easily destroy them, but can energize and light 
them up anew at radio frequencies. Bubbles can rise in the hot 
cluster plasma, perhaps detaching from the parent radio galaxy, yet 
stable against Rayleigh-Taylor and other modes.
\end{abstract}
%\PACS
%52.27Ny,52.55.Lf,52.55.Wq,98.62.En
\section{Introductory Physics}
Many physical configurations begin with a current-carrying flow
propagating through a
surrounding plasma, with an ambient magnetic field. This describes
situations varying from Earthly lightning, to
relativistic electron beams born in diodes and propagating in
meter-long ``drift" tubes, to plasma jets accelerated to high
velocities by black hole accretion disks, eventually erecting
magnetic structures millions of light years long

Generally, primary current flows induce return currents in the
ambient plasma (here assumed, for simplicity, to be initially uniform).
Electric fields driven by induction induce currents proportional to
the rate of change of the primary current, closing the circuit back to the
source. [Fig. 1]  This begins the process of magnetic confinement of jets,
now an accepted view (Benford, 1978). The return current region is
typically much larger than the
primary current's (even for gas-pressure-confined situations, such as
ordinary lightning), and for jets is termed the cocoon. Returning 
currents over larger
cylindrical zones minimizes the kinetic energy cost to the inductive 
field. This follows from the general principle that processes 
minimize total energy in building equilibria. Also, such large 
cocoons increase the inductance, increasing endurance times against 
inductive decay. (Probably reconnection governs decay in such 
equilibria, in the region of torodial field reversal; see below.)
\begin{figure}%[ht]
\begin{center}
\includegraphics[scale=.35]{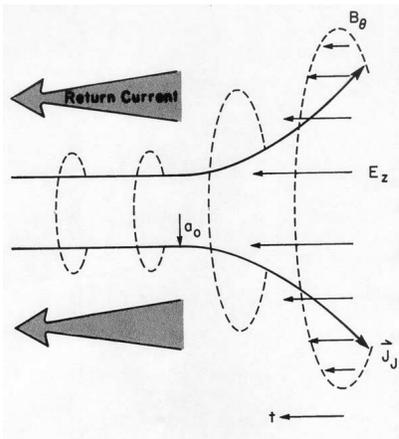}
\end{center}
\caption{A propagating current $J_j$ produces a spatially- and time-varying
magnetic field $B_0$ at its head, driving a return current over an
area much broader than the equilibrium flow radius, $a_0$.  The inductive
$E_z$ drives the return current and time evolution proceeds with
time $t$ increasing behind the head of the flow.}
\end{figure}

Cocoons were invoked from the beginning of astrophysical jet analysis 
(the term is from the late Peter Scheuer). They are large regions
surrounding jets, threaded by magnetic fields and sometimes seen in radiomaps
by synchrotron emission--structures into which the jet has
deposited much of its energy. Current-carrying jets must necessarily
have their return currents carried in this volume, because it is
energetically efficient to induce flows in many particles at low
velocity (versus a jet, which has fewer particles at high velocity). 
This large, mass-loaded cocoon
can preserve confinement of plasma through its self-organized 
magnetic configuration. Beyond the cocoon lies what we
shall call the ``shell," implying that plasma there is not magnetically
confined
(since the polodial field has dropped, perhaps to zero). Its sole
electrodynamic role is to form a
conducting boundary in the sense of Figure 2. Magnetic fields there
can respond passively to the evolution of the jet-cocoon system, and
in environments with pressure gradients can respond to the slow
forces (buoyancy, ram pressures, etc.), independently of the
jet-cocoon structure details.
  Numerical simulations of jet terminations show their hydrodynamical 
nature. Jets with strong toroidal magnetic fields  do not develop 
substantial reverse current cocoons. Instead, the shocked jet plasma 
confined by magnetic stresses forms a ``nose cone"-shaped head. This 
does not describe the cocoons seen in classical FR II radio sources, which 
do not appear to have such nose cones.
This is because
numerical simulations of strongly magnetized jets assume ideal MHD
flows, whereas the current closure condition requires that a large 
fraction of magnetic fields to resistively dissipate, evolving to 
force-free jets (Lesch, Appl, \& Camenzind, 1989; Appl 
and Camenzind, 1992). Plausibly this happens mostly in the very 
compact hot spots observed in radio lobes, but as the Figure 
suggests, jets spread radially and drive inductive return currents 
over large radii, building cocoons much larger than the jets (Lesch 
\& Birk, 1998). This in turn creates reversed field pinches, as the 
structure evolves with immunity to pressure-driven instabilities. 
Like the magnetic well geometries of fusion plasmas, the Taylor state 
equilibria are not merely stable but have a finite margin of 
stability. Even if the kink instability occurs, kinks
seem to lead to helicoidal equilibria with a redistribution
of the current density rather than to disruption (Lery
et al., 2000). Kinks evolve toward the  relaxed Taylor state.
\begin{figure}%[htb]
\begin{center}
\includegraphics[scale=.5]{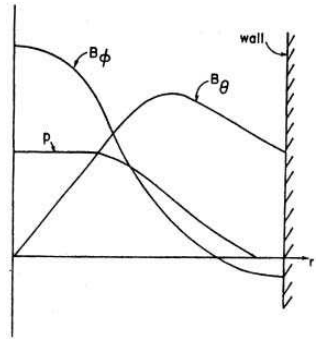}
\end{center}
\caption{The cylindrical reversed field pinch geometry.  In plasma
physics terms the confining polodial field $B_{\theta}$ rises with
radius $r$, while the torodial field $B_{\Phi}$ falls from the flow
axis.  Plasma pressure $p$ is flat throughout most of the region
where it is confined.  A conducting wall defines the boundary
conditions.}
\end{figure}

The physical process by which the accretion disk drives the
jet/lobe is thought to be magnetic helicity injection. Near a black 
hole, differential Keplerian disk rotation
twists up a magnetic arcade above the disk, directly converting 
mechanical energy into magnetic energy.
In the laboratory, a radial electric field imposed by coaxial 
electrodes faithfully simulates the
same process by drawing current from a power supply to run through a 
plasma threaded by an externally
imposed poloidal magnetic field. The toroidal magnetic flux is then injected
by the current into the discharge chamber, driving a magnetic bubble 
expansion. Laboratory electrostatic
helicity injection has found usage from spheromak formation to 
non-inductive current drive in
spherical tori.

	Much analysis has
gone into the shock at the astrophysical jet head, and its injection
of energetic
particles into the cocoon, but often the electrodynamic necessity to return
the current is neglected in equilibria. Yet jets seem often
magnetically self-confined, which implies that cocoons are a major
agency in transmitting external pressures to the jet through the
intermediary magnetic fields.
Laboratory experience echoes this, through of course many of the
dimensionless parameters are very different. In the laboratory, cyclotron
radii are significant, but not usually in astrophysics, except in the
crucial physics of particle acceleration. However, general
stability theorems transcend the differences, within the glade governed by MHD.

	Our hypothesis is that generally these and
similar situations can create
eventually stable, magnetically confined (overall) equilibria that
evolve to a Taylor state . Unstable beams or jets that do
not lead to long-lived magnetic systems do not produce classic radio
sources or well-controlled beams in plasma chambers, and are not
of much interest.

    The usual MHD description of
these systems begins with an assumed
equilibrium (usually very simple) and then considers perturbations affecting
stability. But how to fix the equilibrium? This is harder than linear 
stability theory, since equilibria are fundamentally nonlinear 
solutions of the MHD equations, equating the $j\times B$ force to the 
pressure gradients.

We first argue, from much experience of the plasma physics community,
that a particular class of equilibria, the reversed
field pinch, should emerge.
The reversed field pinch equilibrium in Fig. 2 has high plasma
pressure  and in the laboratory operates with small safety
factors (Bellan, 2000). Both polodial and torodial fields are of similar
magnitude and play important roles in radial pressure balance. A
conducting wall at large radius is often used in laboratory pinches,
and in jets this can be akin to the distant conducting plasma beyond
formation of a cocoon, i.e., beyond the envelope of return current.
The critical signature of the reversed polodial field lends this 
class of equilibria its
name, because in a qualitative sense the contained plasma is
compressed between the outer polodial tension and the inner torodial
pressure (Bellan, 2000).
We conclude, then, that many magnetic relics, sometimes called 
``ghosts" and fossils (En\ss lin, 2003) may well be reversed field 
pinches. This means
their later lives are governed by the magnetic tensions threading such
structures. Particularly, then, it is no mystery that they are stable as they
rise like balloons in the hot plasmas of clusters. Recall that 
ordinary balloons
work because they have a surface tension and stable boundary 
equilibria, responding through their surface tension to outside 
pressures. While external
fields may aid magnetic balloon stability, they are rugged already. 
This can influence much of our thoughts about cluster structures such 
as relics and ghosts (En\ss lin \& Gopal-Krishna, 2001; Jones \& De 
Young,  2005;  Eilek et al, 1999; Feretti, 2000; Giovanninni, et.al, 
1993, 1999; Govoni,F., 2001). Such  MHD-stable bubbles in clusters 
can keep out hot cluster plasmas and if compressed and energized by 
passing shocks can
lead to the observed radio ``ghosts" ( En\ss lin, 1999).

Often in astrophysical models the magnetic field is tangled on scales that
are much smaller than the scales of fluid
motions. Therefore, the plasma of magnetic fields and relativistic
particles is confined to small 'bubbles' intermixed
with the non-relativistic, thermal
plasma. The size of these bubbles is set by the tangling scale of the
magnetic fields. Thus the two fluids are separated
on microscopic scales (see Brunetti,  et. al., 2001,  and references therein).
This picture does not apply to coherently built
magnetic structures, so is antithetical to our approach. (Jones \& De 
Young,  2005;
En\ss lin, 2003)
	Cluster plasmas should have the wide range of values of magnetic field (microGauss), density, coherence lengths of the fields, etc. to allow an ideal MHD approach. Classical (Spitzer) conductivity in the 10 keV plasma yields magnetic diffusion times $\sim$ gigayear over scales of interest. Only in the turbulent setting-up of equilibria by jets should resistivity matter at the working jet head. We thus exploit the advantage of a general stability method over numerical simulations, which depend critically on parameter sizes.

\section{Evolution of Current-Carrying Jets and Their Magnetic Balloons}

	After a time longer than either the rise time of the jet
current, or the Alfven crossing time (whichever is longer),
self-confined magnetic configurations can evolve by the production
of return currents in the ambient plasma, and the interaction
between the primary and return currents. The constellation of ideas
regarding such evolving long-term structures centrally invokes the
concept of magnetic helicity,
\bqe
K = \int_v \vec{A}\cdot\vec{B}\, d^3r
\eqe
here $\va$ is the vector potential of the magnetic field $\vb$.

	 Evolution of magnetic structures built by current-carrying
flows, if they produce magnetically confined beams or astrophysical
jets, should
follow three  concepts developed in the study of laboratory plasma
confinement. The
guiding principles gained from laboratory experience are:
\begin{enumerate}
\item  For time scales less than the resistive diffusion time of the
system, K is conserved. For sizable jets, this can mean essentially
forever, since the diffusion time scales with A, where A  is the
system cross section normal to current flow. (We assume here a
generally cylindrical geometry, with current along the axis.)
\item  The twist of a magnetic field cannot be too large, or it will
be unstable to a variety of modes, particularly the kink. For a
current-flow pattern of size L, instability occurs if
\bqe
\mu_0 I > \psi L
\eqe
where $I$ is the total current along a flux tube $\psi$   and $\mu_0$
is the mks magnetic constant.
\item  $K$ is much better conserved than magnetic energy for
microscopic dissipative processes.
\end{enumerate}

	These principles imply that long-term structures of large
size can evolve by accumulating twist (helicity), and then suffer
disruption that sheds some helicity, returning to a stable state for
a while. Magnetic flux and energy are not conserved during helicity
buildup or shedding. Dissipation of field energy (during
reconnection, principally) can heat plasma (which has pressure p) and
accelerate electrons,
provoking emission of electromagnetic radiation (often, synchrotron).
	Jets and beams can build the long-lived magnetic structures,
following the above three principles. In a sense
these are like leaky thermodynamic systems that have temperature
gradients and non-uniform fluxes. Magnetic systems will have
gradients in the scale parameter $\lambda$  and a non-uniform
helicity flux (Bellan, 2000).

As an astrophysical jet source (presumably a collapsed rotating object with an
accretion disk acting as a dynamo) delivers helicity to the magnetic
volume, little flows back to the source; the two agencies are weakly
coupled. This implies a gradient in  $\lambda$.  Only if there was
little helicity dissipation in the magnetic structure will the
gradient in  $\l$   be small, and so  $\l$    will be nearly uniform.

It seems plausible that the governing, evolved equilibrium of
jet-driven, long-lived magnetic structures will be a reversed field
pinch (Taylor, 1963; Bellan, 2000) . This axisymmetric configuration 
has field components
$B_z\sim B_{\theta}\sim (\mu_0p)\sfr$ in an MHD equilibrium made stable
by optimally efficient radial profiles demanding a minimum of $B_z$.
($B_{\theta}$  is built by the jet current, $I$; Figure 2.)

Stable, high-plasma pressure ($\beta$) reversed field pinch radial
profiles have general several properties:
\begin{enumerate}
\item  A $B_z$ field that reverses near the outside of the
confinement region. This is the crucial shear that stabilizes
interchange modes and prevents formation of kink $(m=1)$
current-driven waves.
\item To suppress ``sausage" modes driven by pressure, a value of
$\beta < \frac{1}{2}$.
\item A conducting ``wall" close enough to the plasma core to
suppress the kink, $m=1$, current-driven internal kink modes. This
also completely damps all external kink modes.
\item A pressure profile $p(r)$ that is hollow or very flat in $r$, to
suppress interchange modes near the magnetic axis.
\end{enumerate}
The fundamental theory describing evolution to a reversed field pinch
state is due to Taylor (Taylor, 1963).
 We envision a (reversed field) pinch
evolving adiabatically through a sequence of minimum-energy states,
as the jet current drives expansion of the structure. Relaxation
occurs behind the head of the jet, where current variations are slow.

Ohmic dissipation alone cannot yield a reversed field
pinch. Some turbulence must maintain reversal by anomalous transport
and plasma convection. Yet the turbulence cannot by definition be so
disruptive as to disallow a long-term stable structure; such cases we
would not see.

Radial pressure balance and overall force balance in the entire
structure, including the return currents, are similar to those for a
simple Z-pinch.  Though the reversed field pinch pressure profile
constraint is significant, these arise naturally in experiments with nearly
force-free Ohmic discharges of high current (Bellan, 2000, p. 354).

The simplest requirement for overall stability against interchange of
flux lines is that the average curvature of the magnetic geometry be
positive.  This is valid for general three-dimensional  closed-line
systems of arbitrary $\beta$, i.e., $p/B^2$.
This is a necessary, and quite general, condition.

We now turn to our major result.  We model the evolution of reversed
field pinch magnetic structures as a ``rigid" (slowly varying, high
inertia) jet, often
relativistic, which by induction drives return currents in the
surrounding plasma, constructing the entire return current structure
and especially the ``cocoon" which immediately surrounds the luminous
jet.  If a ``rigid" (slowly varying) current in the jet generates the magnetic
structure, what stable configuration does it make? And what happens
to the stability of the structure after the jet current ceases, as it must?

\section{Theorem: Stability of Magnetic Well Equilibria with a Rigid Current}

If a jet ebbs, how can we track the stability of the changing 
equilibria? This is a
vast problem and plasma theory has few tools to attack it. Solving 
the time-dependent fluid equations
is hopelessly complicated. Ordinary stability theory identifies the 
failure modes, with growth rates,
but offers little counsel about how the system responds.  One method 
emphasizes ``marginal stability"--
gradual readjustments of gradients or other equilibrium parameters to 
make linear growth rates evolve to zero.
This method has limited use; one still does not know how the 
system adjusts globally.

Here we use a different approach, taking the perfect conductivity 
energy principle analysis to assess stability. For very large 
structures, perfect conductivity is a plausible approximation because 
the scales over which an equilibrium adjusts allows no significant 
role to the diffusion time, even though the jets that set up the 
equilibrium need dissipation to bring about the return current 
circuit that establishes the structure.

This energy principle method sets general
conditions on the equilibrium fields (Johnson et.al, Kulsrud, 1969). 
Rather than tracking individual modes, we think globally about how 
structures
evolve. (Bernstein et. al., 1958; Taylor, 1963) The perfect 
conductivity assumption is essential for energy principle analysis, 
because resistivity implies a steady draining of energy by Ohmic 
dissipation, an extraneous effect outside stability analysis. We 
neglect resistivity because in large structures resistive zones lie 
typically where reconnection proceeds. In reversed field pinches, 
this is usually near where the axial field reverses--i.e., deep in 
the structure, where particles get energized--and so largely beside 
the point of overall stability. The global resistive decay times are 
enormously long, and so negligible.

We cannot follow the jet decay, which would demand a full 
time-dependent analysis. Instead, one can model in snap-shot fashion. 
We take the reversed field pinch in the jet-on state, applying energy 
condition stability criteria, and then compare with the final, 
jet-off state. It will turn out that the conditions on magnetic field 
equilibria are the same. When the jet is on we take
the driver current to be stiff in the MHD sense. This seems appropriate for
jet flows that are either dense or relativistic,
constituting a quasi-rigid current
system, essentially unaffected by the slow reaction of the far larger 
surrounding plasma.  With high kinetic energy,
the jet resists any surrounding magnetic fields. Our strategy is then 
to examine what the powerful energy principle
method says about the long term, once the jet current dies. This seems
plausible because by the time the jet builds magnetic equilibria on 
scales larger than galaxies,
the inductive decay time far exceeds the Hubble time. When the jet 
current dies, the equilibrium adjusts slowly,
so there is no quick perturbing pressure to make it unstable. The 
entire equilibrium adjusts from the jet, radially outward.  End-state 
analysis can frame the
stability issue without following the intricate intermediate evolution. We take two "snapshots" during and after the jet lifetime and find that they have identical stability requirements.

The
effect of the initial current driver should then be included only in the
equilibrium-generating  $\nabla \times
\vec{B}$ equation,
\bqe
\nabla \times \vec{B} = \vec{j}_p+\vec{j}_e ,
\eqe
where $\vec{B} =$ magnetic flux density (in rationalized units, with
$\mu_0 = 1$), $\vec{j}_p =$ plasma current, and $\vec{j}_e =$ driver
current.  The $\vec{j}_e$ term is treated as a fixed function of
position $\vec{j}_e = \vec{j}_e(\vec{x})$.  Apart from this, the
usual magnetohydrodynamic (MHD) equations apply, with
the sole proviso that the $\vec{j}$ appearing in the
$\vec{j}\times\vec{B}$ term of the MHD equation of motion is
$\vec{j}_p$.  Quite generally the usual energy condition holds, with 
some modifications (Bernstein et.al., 1958).

Here we consider the analogous problem for a tensor-pressure
equilibrium of the ``mod-B" type with pressure $p_{\perp}$ perpendicular to the field and $\quad p_{\|}$ parallel to it (Northrop \& Whiteman,
1964)--
  i.e., when both
pressure components are functions of $B (=|\vec{B}|)$:
\bqe
p_{\perp} = p_{\perp}(B),\quad p_{\|} = p_{\|}(B).
\eqe
We assume that all flux lines
ultimately intersect a plasma boundary, and for unbounded cases (such
as jets) this must mean some distant conducting plasma that imposes a
boundary condition on the entire equilibrium.
Here we are concerned with axisymmetric geometries,
the same conclusions follow trivially on
grounds of symmetry.
Such equilibria highly favor containment and stability.

We now show that
this remains true in the presence of an additional $\vec{j}_e$ -- and
therefore that the equilibria built by laboratory beams or
astrophysical jets will have the same stability properties after the
source current turns off.
Specifically, we show that the well-known Hastie-Taylor
criterion applies (Hastie and Taylor, 1964).

Take the equilibrium condition:
\bqe
\vec{j}_p \times \vec{B} &=& \nabla\left[
p_{\perp}+\vec{B}\cdot\nabla
B^{-2}(p_{\|}-p_{\perp})\right]\nonumber\\
&=& (\nabla \times \vec{B})\times \vec{B} -\vec{j}_e\times \vec{B}.
\eqe
Taking the parallel component, one easily finds
\bqe
Bp'_{\|} = p_{\|}-p_{\perp} ,
\eqe
where the ``prime" denotes differentiation with respect to $B$.  In
the case of interest (beam- or jet-generated magnetic structures),
both components are
decreasing functions of $B$, since the magnetic ``well" confines the
plasma. For a reversed field pinch, this confinement occurs in the
broad region near the reversal in sign of the torodial magnetic field
(Fig.2). This generally implies
\bqe
p_{\perp} > p_{\|} .
\eqe

In the absence of a $\vec{j}_e$, it follows  that
\bqe
\vec{j} = - \left( 1 +
\frac{p_{\perp}-p_{\|}}{B^2}\right)^{-1}\left(\frac{p_{\perp}-p_{\|}}{B^2}\right)^{\prime} 
\, \nabla \vec{B}\times \vec{B} ,
\eqe
and from this it follows directly that
\bqe
\vec{j}\cdot\vec{B} = 0.
\eqe
The analogous relations in the present case are
\bqe
&&\left( 1 +
\frac{p_{\perp}-p_{\|}}{B^2}\right)\vec{j}_p+\left(\frac{p_{\perp}-p_{\|}}{B^2}\right)\vec{j}_e\nonumber\\
&& \qquad\qquad = -
\left(\frac{p_{\perp}-p_{\|}}{B^2}\right)^{\prime}\, \nabla
B\times\vec{B},
\eqe
and
\bqe
(B^2 +
p_{\perp}-p_{\|})\vec{j}_p\cdot\vec{B}+(p_{\perp}-p_{\|})\vec{j}_e\cdot\vec{B}
= 0.
\eqe
To derive the stability criterion, we can use the paper of
Taylor and Hastie (1964), making appropriate modifications wherever
necessary.  We note first that their Eq. (4) continues to  apply in
the present case, if $\vec{j}_p$ is substituted for $\vec{j}$ in the
second term.

This can be shown by going through the original
derivation of the guiding-center energy principle (Kruskal and 
Oberman, 1958) and making
sure that whenever using the $\nabla\times\vec{B}$ equation, the
$\vec{j}_e$ term is properly included.
	    In Taylor and Hastie's
subsequent calculations, the $\vec{j}\cdot\vec{B} = 0$ condition was
explicitly used in several places.  If we use Eq. (9) instead,
however, we find that $\vec{j}_e$ terms cancel out  in the end,
thus leading to no change in the final result [their Eq. (1)].

Now we seek general conditions for stability. In its final form, the 
integrand of the energy integral can be
written as $(B+p'_{\perp})$ times a positive term, plus $(B-p'_{\|})$
times a positive term, plus a positive term.  Moreover, by an
appropriate choice of trial functions, the following conditions can
be satisfied simultaneously: (1) the third term is negligible
everywhere.  (2) The first and second terms are negligible everywhere
except in the immediate neighborhood of a single arbitrarily chosen
point $P$.  (3) The ratio of the first two terms in the neighborhood
of $P$ can be made either arbitrarily large or arbitrarily small.
Under these circumstances, it is necessary and sufficient for
stability if both the following conditions are satisfied for all
values of $B$:
\bqe
B-p'_{\|} > 0, \, B+p'_{\perp} > 0.
\eqe
The first condition is always satisfied in the case of interest
(magnetic confinement), and the second sets and upper limit of
$\frac{1}{2} (B_{\rm max}^2-B^2_{\rm min})$ on the maximum of
$p_{\perp}$ at the center of the well. Here $p_{\perp{\rm max}} =
p_{\perp}(B_{\rm min})$; $p_{\perp}(B_{\rm max})=p_{\|}(B_{\rm
max})=0]$.

These are the same as Hastie and Taylor's conclusions,
which have become
standard wisdom in fusion plasma physics. The point is that
they are unaffected by the presence of a nonvanishing
$\vec{j}_e$.
	This implies that a wide range of equilibria available to
propagating currents,
achieved by driving return currents in the larger surrounding plasma,
can be set up and will then persist after the driver current tapers
away (as a beam shuts off, or a black hole jet dies)  because its
stability is ensured.

Laboratory experiments provide much of our lore about reversed field
pinches, and these are
usually low pressure devices (``low beta", where $\beta  = [p_{\perp{\rm
max}}/(B_{\rm max}^2)]$). However, our energy principle stability 
analysis does not
depend upon this critical stability parameter $\beta$ being very
small, only $\beta <1$.  Taylor
argued that in all systems with $\beta <1$ magnetic reconnection 
conserves global helicity, allowing
stable geometries to evolve without losing stability.
	``Unfreezing" the magnetic field lines from plasma demands 
resistivity and thus
some form of magnetic energy dissipation, but helicity can be
preserved during this, as is the case for dissipation by
reconnection.  Thus $\beta \sim 1$ may occur in long-lived systems if
helicity may be shed through turbulent processes other than
reconnection.

\section{Conclusions: the Evolution of Magnetic Bubbles}
We have argued that long-lived magnetic structures generated
by current-carrying flows evolve into Taylor states, and stay that way
for quite long times, probably $>$ a billion years. Our principal 
assertion is that one can model reversed field pinch equilibria by 
taking
the early-age driver current to be stiff in the MHD sense. Stability
conditions of the minimum-B variety, familiar from fusion plasma
studies, apply in vast radio structures while current drivers are on. The
stability condition so reached
will then also obtain for
the later, ``cooling down" state without an active current source,
when magnetic equilibria will persist against
dissipation of magnetic energy.

If these equilibria with jets present cannot shed helicity, K, they can go kink
unstable. After there is no jet, the problem vanishes--kink stability 
can be achieved by expansion
of the equilibrium radius, so that the unstable wavelengths become
longer than the structure length.
Generally, the large magnetic
structures built
by current flows from compact sources can be studied using energy principle
methods and invoking the Taylor logic learned from laboratory cases.
This means that observed MHD-stable bubbles in hot clusters can keep 
out hot cluster plasmas,
leading to radio ``ghosts."
  Stability
conditions assume that magnetic fields exert non-isotropic stresses,
as is critical in flows.  Helical strong fields appear to keep jets
from widening in numerical simulations (Punsley, 2001, En\ss lin et 
al, 1997), and we should
expect this helical field structure for the jet and inner cocoon. The
observed similar axial ratios (width to length) in cocoons of FRII
sources suggests a self-similar evolution, and magnetic confining
structures can satisfy this demand .

  How do these ideas apply to magnetic ``balloons" built by jets? 
The reversed
field pinch is endangered by the current-driven global kink
instability; this seems to be the most likely way for structures to
fail when the current source is on. If a jet  can survives the
current-driven era,  later stability  seems more probable, as
loss of the jet removes a source of free energy.  But suppose the system
fails to shed helicity K as dissipation of magnetic energy proceeds;
this is well known to lead to kink instability.
Recent detections of several ghost cavities in galaxy
clusters (En\ss lin and Heinz,
2002; Soker et al., 2002) -- often, but not always, radioemitting--
suggest that the cluster hot plasma stays separated from the bulk of 
the relativistic plasma  on a
timescale of ~100 Myr. Some leakage of higher energy particles
is not excluded by these observations, of course.

What sort of equilibria are plausible?  Dunn and Fabian (2004) found
limits on k/f, where k is the ratio of the total
relativistic particle energy to that in electrons radiating between
10 MHz to 10 GHz and f is the volume filling factor of the relativistic
plasma. None of their bubbles had a simple equipartition
between the pressures from the relativistic particles and the magnetic
field. Further, k/f had no strong dependence on any
physical parameter of the host cluster, and though at first there seemed to be
two populations -- k/f values around 2, another bunch around 300-this 
did not hold up (Dunn,  Fabian and Taylor, 2005). The apparent 
bimodality of
the k/f distribution could have been explained  as arising from two 
kinds of jets--electron-positron, giving a low value for k, and 
electron-proton. If protons are the extra particles needed to 
maintain pressure
equilibrium, but unseen in the radio emission, k is
high. Also, bimodality
could be caused by
either a non-uniform magnetic field, or a filamentary structure in
the lobes. Both possibilities are consistent with a reversed field 
equilibrium, since fields vary, and especially in the lobes there are 
dissipative processes afoot, which do not smooth out structures. 
Later, thermal plasma  entrained during bubble formation would 
reduce the volume filling factor and provide
extra particles, yielding the calculated values. Variations in 
re-acceleration, which may occur in the field reversal volume from 
reconnection events, could also affect the k/f measure. At this point 
we know too little to infer much. The important lesson is that 
constraints on the magnetic  field obtained by
comparing the synchrotron cooling time to the bubble
age show that no bubbles in the sample are in
equipartition. In a few years
measurements of the rotation measures from sources behind
clusters using EVLA might reveal correlations of the upper limits on k/f
with magnetic field. This could test whether the older a bubble is, then
the larger its value of k/f , from aging
of the relativistic electrons.

Plainly,  stable "balloon" fossils of earlier jets can influence cluster 
evolution by rising as bubbles, conveying energy, and hastening 
vertical mass mixing. 
Rising magnetic balloons can detach from 
their host galaxy by reconnection near their foot points--so they 
typically should be larger, the farther they are from the center. 
This can be checked as a general tendency, once we can resolve many 
such ``ghosts." Heating of clusters can come from the shifting of 
such balloons, allowing gas to infall and warm. Many magnetic 
balloons, small and large, can contribute--rather than, say, one huge 
structure from the central galaxy, which seems energetically 
difficult. Magnetic balloons that resist the incursion of cluster plasma may help explain evolution of the cluster plasma over long times.

\noindent{\bf Acknowledgements}

I thank D. Buote, L. Feretti, G. Giovanninni, P. Bellan and J. Eilek 
for useful discussions.

\end{document}